\documentclass[a4paper]{article}
\usepackage{INTERSPEECH2019}
\usepackage{cite}
\usepackage[OT1]{fontenc}
\usepackage{amsmath,amssymb,amsfonts,mathrsfs,graphicx}
\usepackage{mathrsfs}
\usepackage{algorithm, algorithmic}
\usepackage{enumerate}
\usepackage{multirow}
\usepackage{etoolbox}
\DeclareMathAlphabet\mathbfcal{OMS}{cmsy}{b}{n}

\title{Multichannel Loss Function for Supervised Speech Source Separation \\ by Mask-based Beamforming}
\name{Yoshiki Masuyama$^{1, 2 }$ \thanks{This work was done while Yoshiki Masuyama was an intern at LINE Corporation. Paper accepted on Interspeech 2019.}, Masahito Togami$^2$, Tatsuya Komatsu$^2$}
\address{
  $^1$Department of Intermedia Art and Science, Waseda University, Tokyo, Japan \\
  $^2$LINE Corpolation, Tokyo, Japan}
\email{mas-03151102@akane.waseda.jp, masahito.togami@linecorp.com}

\begin{document}
\setlength{\abovedisplayskip}{3pt} 
\setlength{\belowdisplayskip}{3pt} 
\maketitle
\begin{abstract}
In this paper, we propose two mask-based beamforming methods using a deep neural network (DNN) trained by multichannel loss functions.
Beamforming technique using time-frequency (TF)-masks estimated by a DNN have been applied to many applications where TF-masks are used for estimating spatial covariance matrices.
To train a DNN for mask-based beamforming, loss functions designed for monaural speech enhancement/separation have been employed.
Although such a training criterion is simple, it does not directly correspond to the performance of mask-based beamforming.
To overcome this problem, we use multichannel loss functions which evaluate the estimated spatial covariance matrices based on the multichannel Itakura--Saito divergence.
DNNs trained by the multichannel loss functions can be applied to construct several beamformers.
Experimental results confirmed their effectiveness and robustness to microphone configurations.

\end{abstract}

\section{Introduction}
Speech source separation is a fundamental technique with many applications including automatic speech recognition (ASR) \cite{googlehome, book} and hearing aid \cite{sunohara}.
Although speech source separation with a single microphone is applicable \cite{overview}, that with multiple microphones is more effective because it can take advantage of spatial information \cite{consolidated}.
There exist several unsupervised approaches for multichannel speech source separation including independent component analysis based methods \cite{ica, ica2, yatabe} and local Gaussian model (LGM) based method \cite{duong}.
Meanwhile, motivated by the strong capability of a deep neural network (DNN) to model a spectrogram of a speech, supervised approaches have been paid increasing attention \cite{psmpit, yosioka, chimerabeam, dpclgev}.

In supervised speech source separation, beamforming using a DNN have been mainly studied \cite{psmpit, yosioka, chimerabeam, dpclgev}.
It has also been studied in speech enhancement and noise-robust ASR \cite{heymann, beamnet, ochiai}.
One approach is to estimate the complex-valued filter coefficients by a DNN \cite{fix1, fix2}.
This approach can apply to only the same microphone configurations as in its training.
Another approach is called mask-based beamforming where a TF-mask is used for estimating spatial covariance matrices \cite{heymann}.
After estimating spatial covariance matrices, several beamformers such as minimum variance distortion-less response (MVDR) beamformer \cite{mvdr}, generalized eigenvalue (GEV) beamformer \cite{gev}, and time-invariant multichannel Wiener filter (MWF) \cite{mwf} can be constructed in accordance with applications.
This approach does not depend on microphone configurations, and the effectiveness of the mask-based beamforming has been shown in noise-robust ASR \cite{heymann, beamnet}.

\begin{figure}[t]
\centering
\includegraphics[width=0.99\columnwidth]{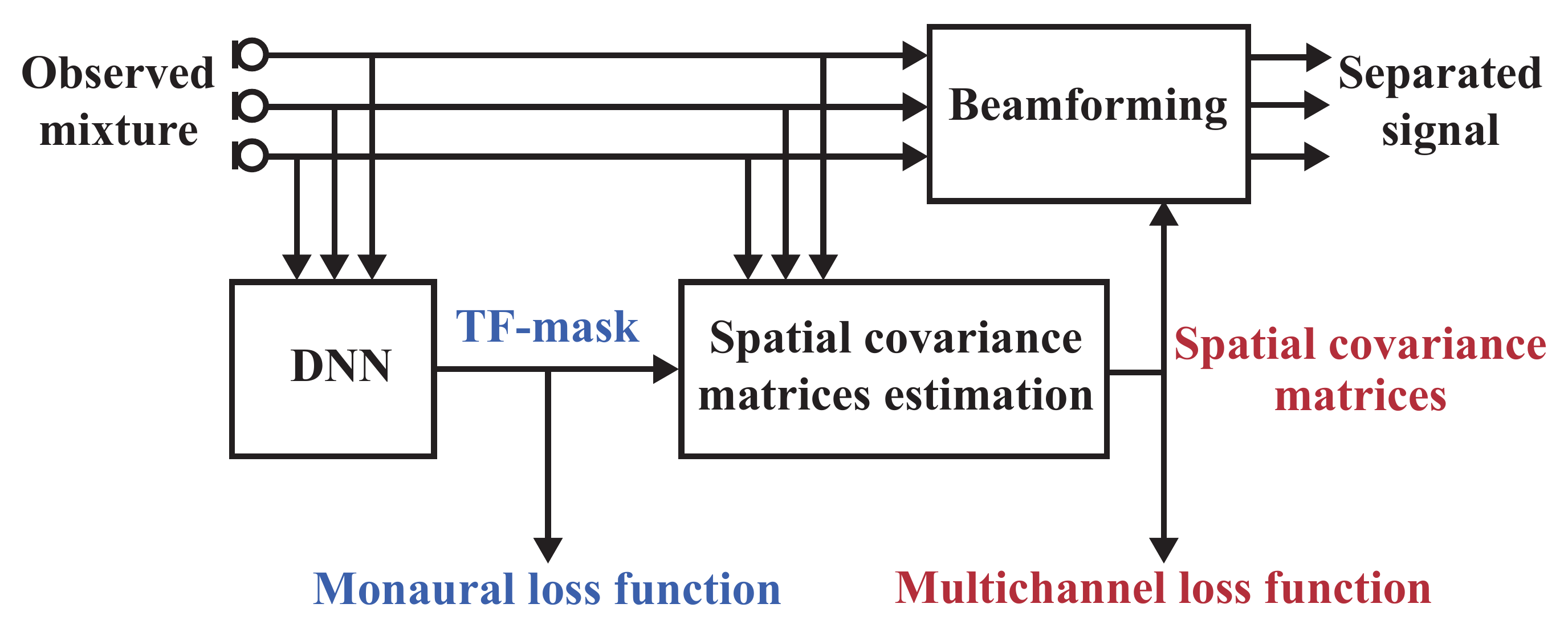}
\vspace{-4pt}
\caption{Block diagram of mask-based beamforming. The proposed methods use multichannel loss functions which evaluate spatial covariance matrices (red) while conventional methods use monaural loss functions (blue).}
\vspace{-6pt}
\label{fig: illust}
\end{figure}

While mask-based beamforming for speech enhancement and noise-robust ASR has been well studied, that for speaker-independent multi-talker separation is still a challenging problem due to the utterance-level permutation problem.
In order to address this issue, permutation invariant training (PIT) was proposed, which solves the permutation problem so that its loss function takes the lowest value \cite{pit, upit}.
In contrast to other approaches to speaker-independent multi-talker separation \cite{dpcl, atract, mdpcl}, PIT can freely design its loss function, and thus the choice of the loss function is important.

Recently, using PIT, speaker-independent multi-talker separation by mask-based beamforming was presented \cite{yosioka, psmpit, chimerabeam}.
In these studies, loss functions designed for monaural speech enhancement/separation, such as the phase sensitive approximation (PSA) \cite{psm}, are employed in the training. 
However, monaural loss functions do not consider inter-microphone information.
A TF-mask considers the signal-to-noise ratio (SNR) at each TF bin, which is not directly related to the spatial covariance matrices.
Meanwhile, the performance of beamforming significantly depends on the estimated spatial covariance matrices.
Hence, the performance of monaural TF-masking does not directly correspond to that of mask-based beamforming.

In this paper, we propose two mask-based beamforming methods with multichannel loss functions.
As illustrated in Fig.~\ref{fig: illust}, the multichannel loss functions evaluate the estimated spatial covariance matrices which are used for constructing beamformers.
A multichannel loss function was originally proposed for the time-varying MWF based on the multichannel Itakura--Saito divergence (MISD) \cite{togami}.
We first import it for time-invariant mask-based beamforming.
Furthermore, since the loss function presented in \cite{togami} is redundant for the time-invariant mask-based beamforming, we also propose the mask-based beamforming with the low-computational loss function.
By using PIT, both proposed methods can be easily applied to speaker-independent multi-talker separation.
Our main contributions are twofold:
(1) proposing mask-based beamforming with multichannel loss functions;
(2) clarifying the effectiveness of multichannel loss functions for several beamformers.

\section{Preriminary}

\subsection{Mask-based beamforming}
Let $N$ source signals be observed by $M$ microphones, $x_{t, f, m}$ be the observed mixture, and $c_{t,f,n,m}$ be the $n$th source signal observed at the $m$th microphone where $t = 1, \ldots, T$ and $f = 1, \ldots, F$ are time and frequency indices, respectively.
A separated source $\hat{c}_{t,f,n}$ obtained by beamforming is given as
\begin{equation}
\hat{c}_{t,f,n} = \mathbf{w}_{f,n}^H \mathbf{x}_{t,f},
\end{equation}
where $\mathbf{w}_{f,n}$ is the time-invariant filter coefficients for extracting $n$th source, and $\mathbf{w}^H$ is the Hermitian transpose of $\mathbf{w}$.
For constructing beamformers, the spatial covariance matrices are required.
Assuming the sparsity of the speeches in TF domain, the spatial covariance matrix of $n$th speech source $\mathbf{R}_{f, n}$ can be estimated as \cite{higuchi}
\begin{equation}
\mathbf{R}_{f, n} = \frac{1}{\sum_{t} \mathcal{M}_{t,f,n}} \sum_{t} \mathcal{M}_{t,f,n} \mathbf{x}_{t, f} \mathbf{x}_{t,f}^{H},
\label{eq: covest}
\end{equation}
where $\mathcal{M}_{t, f, n} \in [0, 1]$ is a TF-mask for extracting the $n$th source, $\mathbf{x}_{t,f} = [x_{t,f,1}, \ldots, x_{t,f,M}]^\top$, and $\mathbf{x}^\top$ is the transpose of $\mathbf{x}$.
Thus, the complex-valued spatial covariance estimation is substituted by the real-valued TF-mask estimation which is independent of the number of microphones.

\subsection{Loss function for TF-mask estimation}
\label{sec: dnnmask}
To train a DNN for TF-mask estimation, several training criteria have been presented such as PSA which minimizes the mean square error between clean and estimated sources on the complex plane.
PSA considers the following loss function:
\begin{equation}
\mathscr{L}_\text{PSA} = \frac{1}{TF} \sum_{t, f} | \mathcal{M}_{t,f,n}{x}_{t, f}-{c}_{t, f, n} |^2,
\label{eq: psm}
\end{equation}
where the microphone index $m$ is omitted because PSA does not requires multichannel observation.
Note that the oracle phase sensitive mask (PSM), achieves the highest SNR in real-valued TF-masking \cite{psm}, and it was recently applied to mask-based beamforming \cite{psmpit}.

However, the performance of monaural speech enhancement/separation does not directly correspond to that of the mask-based beamforming.
This is because such a monaural loss function does not consider inter-microphone information.
Furthermore, TF-mask considers SNR at each TF bin, but it does not directly correspond to the accuracy of the time-invariant spatial covariance matrix calculated by Eq.~\eqref{eq: covest}.

\subsection{Beamformers}
\label{sec: beamformers}
\subsubsection{MVDR beamformer}

MVDR beamformer, which aims to minimize the total power of the extracted source without distortion of the target, is one of the most popular beamformers.
Based on \cite{mvdr}, it is given as
\begin{equation}
\mathbf{w}_{f, n} = \frac{\mathbf{R}_{f, \bar{n}}^{-1}\mathbf{R}_{f, n}\mathbf{e}}{\mathrm{tr}(\mathbf{R}_{f, \bar{n}}^{-1}\mathbf{R}_{f, n})},
\label{eq: mvdr}
\end{equation}
where $\mathbf{R}_{f, n}$ and $\mathbf{R}_{f, \bar{n}}$ are the spatial covariance matrices of the target and interference, respectively, and $\mathbf{e} = [1, 0, \ldots, 0]^\top$.

\subsubsection{GEV beamformer}
GEV beamformer, which aims to maximize SNR for each frequency sub-band, is formulated as \cite{gev}:
\begin{equation}
\mathbf{w}_{f, n} = \text{arg}\max_{\mathbf{w}} \frac{\mathbf{w}^H\mathbf{R}_{f,n}\mathbf{w}}{\mathbf{w}^H\mathbf{R}_{f,\bar{n}} \mathbf{w}}.
\label{eq: gev}
\end{equation}
Note that there exists the ambiguity of complex value scalar multiplication in $\mathbf{w}_{f, n}$.
In \cite{gevgain}, it was solved by minimizing the difference between the estimated source and the observation, which was used in the experiment.

\subsubsection{Multichannel Wiener filter}
Assuming each source signal $\mathbf{c}_{t, f, n}$ independently follows a zero-mean complex-valued Gaussian distribution \cite{duong}:
\begin{align}
\mathbf{c}_{t, f, n} &\sim \mathcal{N}_\mathbb{c} (0, \mathbfcal{R}_{t, f, n}), \\
\mathbfcal{R}_{t, f, n} &= v_{t, f, n}\mathbf{R}_{f, n},
\end{align}
where $v_{t, f, n} \in \mathbb{R}_+$ is the time-varying activation of the $n$th source,
the observed mixture $\mathbf{x}_{t,f,m} = \sum_{n} c_{t,f,n,m}$ follows
\begin{equation}
\mathbf{x}_{t, f} \sim \mathcal{N}_\mathbb{c} (0, \sum_n\mathbfcal{R}_{t, f, n}).
\end{equation}
Then, time-varying MWF can be obtained in the minimum mean square error sense as
\begin{equation}
\mathbf{W}_{t, f, n} = \mathbfcal{R}_{t, f, n}\bigl(\sum_{l}\mathbfcal{R}_{t, f, l}\bigr)^{-1}.
\label{eq: tMWF}
\end{equation}
While Eq.~\eqref{eq: tMWF} is a time-varying filter, its time-invariant version can calculate replacing $\mathbfcal{R}_{t, f, n}$ by $\mathbf{R}_{f, n}$ \cite{mwf, mwf2}.

\section{Proposed mask-based beamforming with multichannel loss function}

In this paper, we propose two mask-based beamforming methods using DNNs trained by multichannel loss functions which evaluate the estimated spatial covariance matrices as illustrated in Fig.~\ref{fig: illust}.
After reviewing a multichannel loss function for time-varying MWF \cite{togami}, the proposed time-invariant mask-based beamforming is introduced, which is based on the same loss function used in \cite{togami}.
Since the loss function focuses on the time-varying MWF, it requires the estimated time-varying activation which is redundant for time-invariant beamforming.
Hence, we also propose a mask-based beamforming method based on another loss function which does not require the estimation of the time-varying activation.

\subsection{Multichannel loss function for time-varying MWF \cite{togami}}
For time-varying MWF, we proposed a multichannel loss function which evaluates the estimated time-varying spatial covariance matrices $\hat{\mathbfcal{R}}_{t, f, n}$.
In \cite{togami}, a DNN estimates the time-varying activation and TF-mask.
Based on DNN's outputs, the time-varying spatial covariance matrices are calculated as $\hat{\mathbfcal{R}}_{t, f, n} = \hat{v}_{t, f, n} \hat{\mathbf{R}}_{f, n}$ where $\hat{\mathbf{R}}_{f, n}$ is given by Eq.~\eqref{eq: covest}.
Then, the loss function based on the MISD \cite{mnmf} between the clean source signal $\mathbf{c}_{t,f,n}$ and estimated one $\hat{\mathbf{c}}_{t,f,n}$ is given by
\begin{align}
\mathscr{L}_1 &= \sum_{t,f,n}
\mathbf{d}_{t,f,n}^H \boldsymbol{\Psi}_{t,f,n}^{-1} \mathbf{d}_{t,f,n} + \log \text{det}(\boldsymbol{\Psi}_{t,f,n}), \label{eq: pmisd} \\
\mathbf{d}_{t,f,n} &= \mathbf{c}_{t,f,n} - \hat{\mathbf{c}}_{t,f,n}, \\
\hat{\mathbf{c}}_{t,f,n} &= \mathbf{W}_{t, f, n} \mathbf{x}_{t,f}, \label{eq: evalmwf} \\
\boldsymbol{\Psi}_{t,f,n} &= (\mathbf{I} - \mathbf{W}_{t, f, n})\hat{\mathbfcal{R}}_{t, f, n},
\end{align}
where $\mathbf{I} \in \mathbb{R}^{M \times M}$ is the identity matrix, and time-varying MWF is calculated as in Eq.~\eqref{eq: tMWF}.
Note that the multichannel loss function given in Eq.~\eqref{eq: pmisd} corresponds to the negative log-likelihood of the posterior distribution $p(\mathbf{c}_{t,f,n} | \mathbf{x}_{t,f})$.

\subsection{Mask-based beamforming with multichannel loss function given in Eq.~\eqref{eq: pmisd}}

The effectiveness of the multichannel loss function given in Eq.~\eqref{eq: pmisd} was confirmed for time-varying MWF \cite{togami}.
As a time-invariant version of  \cite{togami}, we propose a mask-based beamforming method based on the multichannel loss function given in Eq.~\eqref{eq: pmisd}.
Specifically, the proposed method uses the same DNN as in \cite{togami} where the DNN estimates both time-varying activation and time-invariant spatial covariance matrices in its training.
In the testing phase, the DNN estimates only time-invariant spatial covariance matrices for constructing several time-invariant beamformers.

In conventional mask-based beamforming, a DNN is trained to maximize the performance of monaural speech enhancement/separation.
In contrast, the proposed approach trains a DNN based on the model of multichannel signal \cite{duong}, and TF-masks are trained to estimate the accurate spatial covariance matrices.
The effectiveness of this approach was confirmed in experiments in Section~\ref{sec: experiment} where it is referred to as Prop.~$1$.

\subsection{Mask-based beamforming with low-computational multichannel loss function}

In the aforementioned method, a DNN estimates both time-varying activation and spatial covariance matrices, but the estimation of the time-varying activation is redundant for mask-based beamforming because it is not used for constructing time-invariant beamformers.
In addition, minimizing the loss function given in Eq.~\eqref{eq: pmisd} requires huge computation for estimating clean source $\hat{\mathbf{c}}_{t,f,n}$ by time-varying MWF.

In order to address these problems, we propose a mask-based beamforming method using another multichannel loss function given by
\begin{align}
\mathscr{L}_2 &= \sum_{t,f}
\mathrm{tr} (\mathbf{X}_{t,f} \hat{\mathbf{X}}_{t,f}^{-1}) + \log\text{det} (\hat{\mathbf{X}}_{t,f}), \label{eq: misd}\\
\hat{\mathbf{X}}_{t,f} &= \sum_n v^\star_{t,f,n} \hat{\mathbf{R}}_{f, n},
\end{align}
where $\mathbf{X}_{t,f} = \mathbf{x}_{t,f} \mathbf{x}_{t,f}^H$, $\hat{\mathbf{R}}_{f, n}$ is calculated by Eq.~\eqref{eq: covest} as in mask-based beamforming, and $v^\star_{t,f,n}$ is the time-varying activation calculated from the oracle multichannel signal as
\begin{equation}
v^\star_{t,f,n} = \frac{1}{M} \sum_m  \frac{ |c_{t, f, n, m}|^2}{
\sum_t |c_{t, f, n, m}|^2/T},
\end{equation}
which represents the fluctuation from the average power for each source.
While the loss function given in Eq.~\eqref{eq: pmisd} considers the estimated clean source, that in Eq.~\eqref{eq: misd} corresponds to the MISD between $\mathbf{X}_{t,f}$ and $\hat{\mathbf{X}}_{t,f}$, which corresponds to the maximum likelihood estimation for $p(\mathbf{x}_{t,f})$ \cite{mnmf}.
The proposed loss function given in Eq.~\eqref{eq: misd} requires less computation comparing to that in Eq.~\eqref{eq: pmisd} thanks to avoiding time-varying MWF calculation.
In addition, by avoiding the estimation of the time-varying activation, the redundant DNN parameters for mask-based beamforming are eliminated.
This approach will be referred to as Prop.~$2$ in the experiment.

When applying speaker-independent multi-talker separation, there exists the permutation problem between the estimated spatial covariance matrices and the oracle time-varying activation.
In order to solve this problem, we can use PIT \cite{upit}.
That is, the permutation problem is solved so that the loss function takes small value.

\section{Experiment}
\label{sec: experiment}

In order to confirm the effectiveness of the multichannel loss functions, DNNs trained by PSA in Eq.~\eqref{eq: psm} \cite{psmpit} and by multichannel loss functions were compared in speaker-independent multi-talker separation by the mask-based beamforming.
Based on the spatial covariance matrices estimated by TF-masking, three beamformers (MVDR beamformer in Eq.~\eqref{eq: mvdr}, GEV beamformer in Eq.~\eqref{eq: gev}, and time-invariant MWF) were tested.
In addition, we also evaluated \cite{togami} which uses time-varying MWF and mask-based beamformers with oracle PSM.
{\renewcommand\arraystretch{1.1}
\begin{table}[t!]
\centering
\footnotesize
\caption{Details of datasets}
\vspace{-10pt}
\label{tab: conditions}
\begin{tabular}{c|ccc}
\hline \hline
& Mic arrangement [cm] & Corpus & $\text{RT}_{60}$ [ms] \\
\hline
\multirow{2}{*}{Train} & $3$-$3$-$3$-$8$-$3$-$3$-$3$ & \multirow{2}{*}{Train} & \multirow{2}{*}{$160$} \\
      & $8$-$8$-$8$-$8$-$8$-$8$-$8$ &       &       \\
\hline
\multirow{2}{*}{Condition~$1$} & $3$-$3$-$3$-$8$-$3$-$3$-$3$ & \multirow{2}{*}{Test} & \multirow{2}{*}{$160$} \\
      & $8$-$8$-$8$-$8$-$8$-$8$-$8$ &       &       \\
Condition~$2$ & $4$-$4$-$4$-$8$-$4$-$4$-$4$ & Test & $160$ \\
Condition~$3$ & $4$-$4$-$4$-$8$-$4$-$4$-$4$ & Test & $360$ \\
\hline
\end{tabular}
\end{table}
}

\begin{figure}[t!]
\centering
\includegraphics[width=0.99\columnwidth]{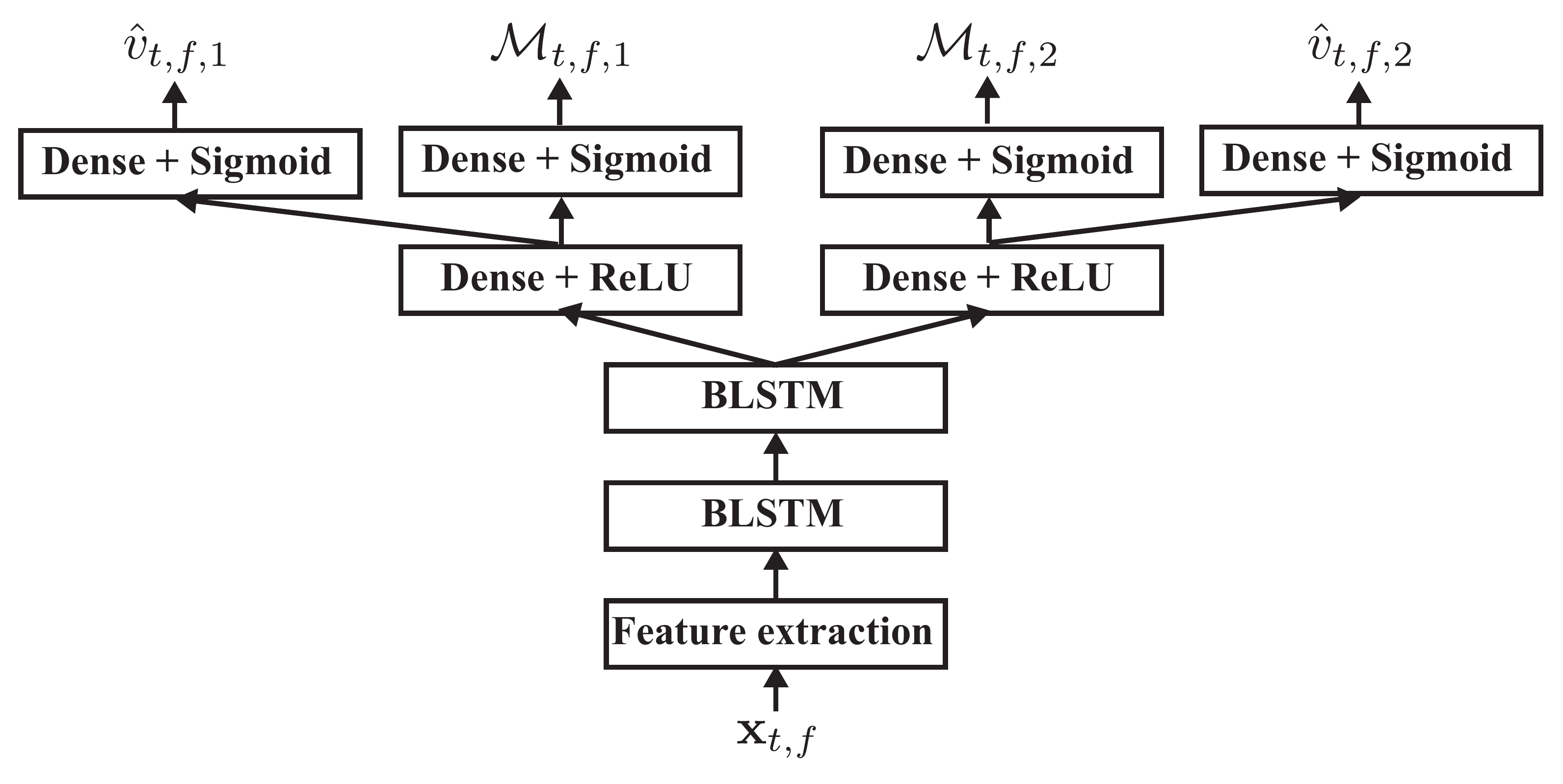}
\vspace{-4pt}
\caption{Network architecture used in experiment.
Mask-based beamformers were calculated from TF-masks $\mathcal{M}_{t,f,n}$.
Time-varying activation $v_{t,f,n}$ was used in only Prop.~$1$ and \cite{togami} for calculating time-varying spatial covariance matrices.}
\vspace{-6pt}
\label{fig: net}
\end{figure}

\subsection{Experimental conditions}
{\renewcommand\arraystretch{1.1}
\begin{table*}[t]
\centering
\footnotesize
\caption{Results of speech separation in Condition~$1$ (closed mic-arrangement for $\text{RT}_{60}$ of $160$ [ms]).}
\vspace{-10pt}
\label{tab: result1}
\begin{tabular}{c|ccc||ccc||ccc}
\hline \hline
& \multicolumn{3}{|c||}{MVDR beamformer}&\multicolumn{3}{c||}{GEV beamformer} & \multicolumn{3}{c}{MWF}\\ 
\hline \hline
Approaches& SDR [dB] & SIR [dB] & CD [dB] & SDR [dB] & SIR [dB] & CD [dB] & SDR [dB] & SIR [dB] & CD [dB] \\ \hline
Mixed  & 0.20 & 0.91  & 4.44 & - & -  & - & - & - & - \\ \hline
PSA \cite{psmpit} & 6.87 & 7.63  & 3.68 & 7.57 & 9.12  & 3.44 & 5.79 & 6.21 & 3.93 \\ \hline
Prop. $1$ & \textbf{8.54} & \textbf{9.42} & \textbf{3.14} & \textbf{8.35} & \textbf{9.76} & \textbf{3.13} & \textbf{7.10} & \textbf{7.57} & \textbf{3.40} \\ \hline
Prop. $2$ & 7.86 & 8.92 & 3.25 & 7.83 & 9.35 & 3.16 & 6.76 & 7.37 & 3.56 \\ \hline \hline 
Time-varying \cite{togami}& - & - & - & - & - & - & 8.69 & 11.56 & 3.11 \\ \hline
Oracle PSM &  10.75 & 11.87 & 2.84 & 10.75 & 12.15 & 2.82 & 10.43 & 11.04 & 3.09 \\ \hline
\end{tabular}
\vspace{8pt}
\caption{Results of speech separation in Condition~$2$ (open mic-arrangement for $\text{RT}_{60}$ of $160$ [ms]).}
\vspace{-10pt}
\label{tab: result2}
\begin{tabular}{c|ccc||ccc||ccc}
\hline \hline
& \multicolumn{3}{|c||}{MVDR beamformer}&\multicolumn{3}{c||}{GEV beamformer} & \multicolumn{3}{c}{MWF}\\ 
\hline \hline
Approaches& SDR [dB] & SIR [dB] & CD [dB] & SDR [dB] & SIR [dB] & CD [dB] & SDR [dB] & SIR [dB] & CD [dB] \\ \hline
Mixed  & 0.20 & 0.86  & 4.48 & - & -  & - & - & - & - \\ \hline
PSA \cite{psmpit} & 6.31 & 6.93  & 3.78 & 6.82 & 8.26  & 3.58 & 5.40 & 5.80 & 4.00 \\ \hline
Prop. $1$ & \textbf{7.88} & \textbf{8.62} & \textbf{3.27} & \textbf{7.72} & \textbf{8.98} & \textbf{3.24} & \textbf{6.42} & \textbf{6.93} & \textbf{3.53} \\ \hline
Prop. $2$ & 7.05 & 7.94 & 3.43 & 7.07 & 8.45 & 3.37 & 6.17 & 6.78 & 3.69 \\ \hline \hline 
Time-varying \cite{togami}& - & - & - & - & - & - & 7.75 & 10.49 & 3.24 \\ \hline 
Oracle PSM & 10.52 & 11.47 & 2.96 & 10.47 & 11.74 & 2.94 & 10.36 & 10.98 & 3.18 \\ \hline
\end{tabular}
\vspace{8pt}
\caption{Results of speech separation in Condition~$3$ (open mic-arrangement for $\text{RT}_{60}$ of $360$ [ms]).}
\vspace{-10pt}
\label{tab: result3}
\begin{tabular}{c|ccc||ccc||ccc}
\hline \hline
& \multicolumn{3}{|c||}{MVDR beamformer}&\multicolumn{3}{c||}{GEV beamformer} & \multicolumn{3}{c}{MWF}\\ 
\hline \hline
Approaches& SDR [dB] & SIR [dB] & CD [dB] & SDR [dB] & SIR [dB] & CD [dB] & SDR [dB] & SIR [dB] & CD [dB] \\ \hline
Mixed  & 0.18 & 0.90  & 4.05 & - & -  & - & - & - & - \\ \hline
PSA \cite{psmpit} & 3.69 & 4.32  & 3.74 & 3.82 & 5.54  & 3.68 & 3.68 & 4.04 & 3.84 \\ \hline
Prop. $1$ & \textbf{4.59} & \textbf{5.56} & \textbf{3.49} & \textbf{4.40} & \textbf{6.08} & \textbf{3.49} & \textbf{4.26} & 4.77 & \textbf{3.60} \\ \hline
Prop. $2$ & 4.09 & 4.94 & 3.56 & 4.02 & 5.64 & 3.54 & \textbf{4.26} & \textbf{4.79} & 3.68 \\ \hline \hline 
Time-varying \cite{togami}& - & - & - & - & - & - & 5.91 & 8.29 & 3.35 \\ \hline
Oracle PSM & 6.45 & 7.80 & 3.26 & 6.32 & 8.30 & 3.25 & 7.21 & 7.94 & 3.38 \\ \hline
\end{tabular}
\vspace{-8pt}
\end{table*}
}

\subsubsection{Datasets}
In both training and testing phases, the measured impulse response in Multichannel Impulse Response Database (MIRD) \cite{MIRD} and the clean speech in TIMIT corpus \cite{TIMIT} were used for making multichannel signals.
The training and $3$ testing conditions are summarized in Table~\ref{tab: conditions}.
The number of microphones and sources were set to $2$ where $2$ microphones were randomly selected for each sample from the microphone arrangement shown in Table~\ref{tab: conditions}.
For training, $20000$ speeches were selected, and they were split into every $100$ frames in TF domain.
While Condition~$1$ used the same microphone arrangement as training in the testing phase, Condition~$2$ employed different microphone array.
In Condition~$3$, performances in longer reverberation case were evaluated.
In all conditions, the distance between speech sources and microphones was set to $1$ m, and the azimuth of each talker is randomly selected for each sample.
All the speeches were resampled at $8$ kHz, and the short-time Fourier transform was computed using the Hann window whose length was $32$ ms with $8$ ms shift.

\subsubsection{DNN architecture and training setup}
A DNN used in this experiment is illustrated in Fig.~\ref{fig: net}, which  contains two bidirectional long-short term memory (BLSTM) layers, each with $300$ units in each direction, followed by $2$ parallel dense layers where the estimation of the time-varying activation was used for only Prop.~$1$ and \cite{togami}.
Dropout of $0.3$ was applied to the output of each BLSTM.
In all methods, input feature was calculated by
\begin{equation}
\Phi_{t, f} = \mathscr{P}\left[\log \Bigl(\frac{1}{M} \sum_m |x_{t, f, m}|\Bigr)\right],
\end{equation}
where $\mathscr{P}$ is the utterance-level mean and variance normalization.
DNN parameters were updated $10000$ times where the batchsize is $128$, the Adam optimizer was used, and the learning rate was $0.001$.

\subsection{Experimental results}

The performance of speech source separation was evaluated by the signal-to-distortion ratio (SDR) and signal-to-interference ratio (SIR) from BSS-EVAL \cite{bss}, and cepstrum distortion (CD).
The separation results are summarized in Tables~\ref{tab: result1}--\ref{tab: result3} where the scores of the unprocessed mixed signal are omitted in GEV beamformer and MWF because they are the same as in MVDR beamformer.
Prop.~$1$ achieved the highest scores in mask-based beamforming, and Prop.~$2$ also resulted in better scores than PSM.
In addition, MVDR beamformer with Prop.~$1$ achieved comparable SDR and CD with time-varying MWF when $\text{RT}_{60}$ is $160$ [ms].
That is, the multichannel loss functions can be applied to not only the time-varying MWF but also several time-invariant beamformers.
We stress that MVDR beamformer is more preferred in many applications such as ASR because it does not cause artificial noise.
Comparing Tables~\ref{tab: result1} and \ref{tab: result2}, both proposed methods with multichannel loss functions worked well even if the microphone arrangement is different from training.
That is, they can be applied to mask-based beamforming in different microphone arrangements as the conventional monaural losses.
Furthermore, they also worked with longer reverberation as illustrated in Table~\ref{tab: result3}.

Prop.~$1$ achieved better scores than Prop.~$2$ in most cases.
That is, the joint estimation of the spatial covariance matrices with the time-varying activation improved the quality of the estimated spatial covariance matrices where the joint estimation can be interpreted as multi-task training.
However, training of Prop.~$1$ takes $3.1$ times slower than that of Prop.~$2$ with "NVIDIA Tesla V100" because Prop.~$1$ requires the calculation of time-varying MWF as in Eq.~\eqref{eq: evalmwf}.

\section{Conclusion}
In this paper, we proposed two mask-based beamforming methods using DNNs trained by multichannel loss functions.
Two multichannel loss functions, used in the proposed methods, evaluate the spatial covariance matrices based on two types of MISD.
The experimental results indicate that the mask-based beamforming with the multichannel loss functions outperformed that with the monaural loss function regardless of the microphone arrangements.
Hence, we conclude the multichannel loss function is effective for various mask-based beamforming techniques.

\section{Acknowledgements}
The authors would like to thank Dr. Kohei Yatabe for his valuable comments and discussion.

\clearpage

\bibliographystyle{IEEEtran}

\begin{thebibliography}{10}

\bibitem{googlehome}
B.~Li, T.~N. Sainath, A.~Narayanan, J.~Caroselli, M.~Bacchiani, A.~Misra,
  I.~Shafran, H.~Sak, G.~Pundak, K.~Chin, K.~C. Sim, R.~J. Weiss, K.~W. Wilson,
  E.~Variani, C.~Kim, O.~Siohan, M.~Weintraub, E.~McDermott, R.~Rose, and
  M.~Shannon, ``Acoustic modeling for google home,'' in \emph{Proc.
  Interspeech}, Aug. 2017, pp. 399--403.

\bibitem{book}
S.~Watanabe, M.~Delcroix, F.~Metze, and J.~R. Hershey, Eds., \emph{New Era for
  Robust Speech Recognition: {E}xploiting Deep Learning}.\hskip 1em plus 0.5em
  minus 0.4em\relax Springer, 2017.

\bibitem{sunohara}
M.~{Sunohara}, C.~{Haruta}, and N.~{Ono}, ``Low-latency real-time blind source
  separation for hearing aids based on time-domain implementation of online
  independent vector analysis with truncation of non-causal components,'' in
  \emph{IEEE Int. Conf. on Acoust., Speech Signal Process. (ICASSP)}, Mar.
  2017, pp. 216--220.

\bibitem{overview}
D.~{Wang} and J.~{Chen}, ``Supervised speech separation based on deep learning:
  {A}n overview,'' \emph{IEEE/ACM Trans. Audio, Speech, Lang. Process.},
  vol.~26, no.~10, pp. 1702--1726, Oct. 2018.

\bibitem{consolidated}
S.~Gannot, E.~Vincent, S.~Markovich-Golan, and A.~Ozerov, ``A consolidated
  perspective on multimicrophone speech enhancement and source separation,''
  \emph{IEEE/ACM Trans. Audio, Speech and Lang. Proc.}, vol.~25, no.~4, pp.
  692--730, Apr. 2017.

\bibitem{ica}
P.~Smaragdis, ``Blind separation of convolved mixtures in the frequency
  domain,'' \emph{Neurocomputing}, vol.~22, no.~1, pp. 21--34, 1998.

\bibitem{ica2}
H.~{Saruwatari}, T.~{Kawamura}, T.~{Nishikawa}, A.~{Lee}, and K.~{Shikano},
  ``Blind source separation based on a fast-convergence algorithm combining
  {I}{C}{A} and beamforming,'' \emph{IEEE Trans. Audio, Speech, Lang.
  Process.}, vol.~14, no.~2, pp. 666--678, Mar. 2006.

\bibitem{yatabe}
K.~Yatabe and D.~Kitamura, ``Determined blind source separation via proximal
  splitting algorithm,'' in \emph{IEEE Int. Conf. Acoust., Speech, Signal
  Process. (ICASSP)}, Apr. 2018, pp. 776--780.

\bibitem{duong}
N.~Q.~K. {Duong}, E.~{Vincent}, and R.~{Gribonval}, ``Under-determined
  reverberant audio source separation using a full-rank spatial covariance
  model,'' \emph{IEEE Trans Audio, Speech, Lang. Process.}, vol.~18, no.~7, pp.
  1830--1840, Sep. 2010.

\bibitem{psmpit}
L.~Yin, Z.~Wang, R.~Xia, J.~Li, and Y.~Yan, ``Multi-talker speech separation
  based on permutation invariant training and beamforming,'' in
  \emph{Interspeech}, Sep. 2018, pp. 851--855.

\bibitem{yosioka}
T.~{Yoshioka}, H.~{Erdogan}, Z.~{Chen}, and F.~{Alleva}, ``Multi-microphone
  neural speech separation for far-field multi-talker speech recognition,'' in
  \emph{IEEE Int. Conf. on Acoust., Speech Signal Process. (ICASSP)}, Apr.
  2018, pp. 5739--5743.

\bibitem{chimerabeam}
Z.~{Wang} and D.~{Wang}, ``Combining spectral and spatial features for deep
  learning based blind speaker separation,'' \emph{IEEE/ACM Trans. Audio,
  Speech, Lang. Process.}, vol.~27, no.~2, pp. 457--468, Feb. 2019.

\bibitem{dpclgev}
L.~Drude and R.~Haeb-Umbach, ``Tight integration of spatial and spectral
  features for {B}{S}{S} with deep clustering embeddings,'' in
  \emph{Interspeech}, Aug. 2017, pp. 2650--2654.

\bibitem{heymann}
J.~{Heymann}, L.~{Drude}, and R.~{Haeb-Umbach}, ``Neural network based spectral
  mask estimation for acoustic beamforming,'' in \emph{IEEE Int. Conf. on
  Acoust., Speech Signal Process. (ICASSP)}, Mar. 2016, pp. 196--200.

\bibitem{beamnet}
J.~{Heymann}, L.~{Drude}, C.~{Boeddeker}, P.~{Hanebrink}, and R.~{Haeb-Umbach},
  ``Beamnet: End-to-end training of a beamformer-supported multi-channel
  {A}{S}{R} system,'' in \emph{IEEE Int. Conf. on Acoust., Speech Signal
  Process. (ICASSP)}, 2017.

\bibitem{ochiai}
T.~{Ochiai}, S.~{Watanabe}, T.~{Hori}, J.~R. {Hershey}, and X.~{Xiao},
  ``Unified architecture for multichannel end-to-end speech recognition with
  neural beamforming,'' \emph{IEEE J. Selected Topics Signal Process.},
  vol.~11, no.~8, pp. 1274--1288, Dec. 2017.

\bibitem{fix1}
B.~Li, T.~N. Sainath, R.~J. Weiss, K.~W. Wilson, and M.~Bacchiani, ``Neural
  network adaptive beamforming for robust multichannel speech recognition,'' in
  \emph{Interspeech}, 2016, pp. 1976--1980.

\bibitem{fix2}
X.~{Xiao}, S.~{Watanabe}, H.~{Erdogan}, L.~{Lu}, J.~{Hershey}, M.~L. {Seltzer},
  G.~{Chen}, Y.~{Zhang}, M.~{Mandel}, and D.~{Yu}, ``Deep beamforming networks
  for multi-channel speech recognition,'' in \emph{IEEE Int. Conf. on Acoust.,
  Speech Signal Process. (ICASSP)}, 2016, pp. 5745--5749.

\bibitem{mvdr}
M.~{Souden}, J.~{Benesty}, and S.~{Affes}, ``On optimal frequency-domain
  multichannel linear filtering for noise reduction,'' \emph{IEEE Trans. Audio,
  Speech, Lang. Process.}, vol.~18, no.~2, pp. 260--276, Feb. 2010.

\bibitem{gev}
E.~{Warsitz} and R.~{Haeb-Umbach}, ``Blind acoustic beamforming based on
  generalized eigenvalue decomposition,'' \emph{IEEE Trans. Audio, Speech,
  Language Process}, vol.~15, no.~5, pp. 1529--1539, 2007.

\bibitem{mwf}
S.~{Doclo} and M.~{Moonen}, ``{G}{S}{V}{D}-based optimal filtering for single
  and multimicrophone speech enhancement,'' \emph{IEEE Trans. Signal Process.},
  vol.~50, no.~9, pp. 2230--2244, Sep. 2002.

\bibitem{pit}
D.~{Yu}, M.~{Kolbæk}, Z.~{Tan}, and J.~{Jensen}, ``Permutation invariant
  training of deep models for speaker-independent multi-talker speech
  separation,'' in \emph{IEEE Int. Conf. on Acoust., Speech Signal Process.
  (ICASSP)}, Mar. 2017, pp. 241--245.

\bibitem{upit}
M.~Kolb{\ae}k, D.~Yu, Z.-H. Tan, and J.~Jensen, ``Multitalker speech separation
  with utterance-level permutation invariant training of deep recurrent neural
  networks,'' \emph{IEEE/ACM Trans. Audio, Speech Lang. Proc.}, vol.~25,
  no.~10, pp. 1901--1913, Oct. 2017.

\bibitem{dpcl}
J.~R. {Hershey}, Z.~{Chen}, J.~{Le Roux}, and S.~{Watanabe}, ``Deep clustering:
  {D}iscriminative embeddings for segmentation and separation,'' in \emph{IEEE
  Int. Conf. on Acoust., Speech Signal Process. (ICASSP)}, Mar. 2016, pp.
  31--35.

\bibitem{atract}
Z.~{Chen}, Y.~{Luo}, and N.~{Mesgarani}, ``Deep attractor network for
  single-microphone speaker separation,'' in \emph{IEEE Int. Conf. on Acoust.,
  Speech Signal Process. (ICASSP)}, Mar. 2017, pp. 246--250.

\bibitem{mdpcl}
Z.~{Wang}, J.~{Le Roux}, and J.~R. {Hershey}, ``Multi-channel deep clustering:
  Discriminative spectral and spatial embeddings for speaker-independent speech
  separation,'' in \emph{IEEE Int. Conf. on Acoust., Speech Signal Process.
  (ICASSP)}, Apr. 2018, pp. 1--5.

\bibitem{psm}
H.~{Erdogan}, J.~R. {Hershey}, S.~{Watanabe}, and J.~{Le Roux},
  ``Phase-sensitive and recognition-boosted speech separation using deep
  recurrent neural networks,'' in \emph{IEEE Int. Conf. on Acoust., Speech
  Signal Process. (ICASSP)}, Apr. 2015, pp. 708--712.

\bibitem{togami}
M.~{Togami}, ``Multi-channel itakura saito distance minimization with deep
  neural network,'' in \emph{IEEE Int. Conf. on Acoust., Speech Signal Process.
  (ICASSP)}, May 2019.

\bibitem{higuchi}
T.~{Higuchi}, K.~{Kinoshita}, N.~{Ito}, S.~{Karita}, and T.~{Nakatani},
  ``Frame-by-frame closed-form update for mask-based adaptive {M}{V}{D}{R}
  beamforming,'' in \emph{IEEE Int. Conf. on Acoust., Speech Signal Process.
  (ICASSP)}, 2018, pp. 531--535.

\bibitem{gevgain}
S.~{Araki}, H.~{Sawada}, and S.~{Makino}, ``Blind speech separation in a
  meeting situation with maximum {S}{N}{R} beamformers,'' in \emph{IEEE Int.
  Conf. on Acoust., Speech Signal Process. (ICASSP)}, vol.~1, Apr. 2007, pp.
  41--44.

\bibitem{mwf2}
S.~{Sivasankaran}, A.~A. {Nugraha}, E.~{Vincent}, J.~A. {Morales-Cordovilla},
  S.~{Dalmia}, I.~{Illina}, and A.~{Liutkus}, ``Robust {A}{S}{R} using neural
  network based speech enhancement and feature simulation,'' in \emph{IEEE
  Workshop Autom. Speech Recognit. Underst. (ASRU)}, Dec. 2015, pp. 482--489.

\bibitem{mnmf}
H.~{Sawada}, H.~{Kameoka}, S.~{Araki}, and N.~{Ueda}, ``Multichannel extensions
  of non-negative matrix factorization with complex-valued data,'' \emph{IEEE
  Trans. Audio, Speech, Lang. Process.}, vol.~21, no.~5, pp. 971--982, May
  2013.

\bibitem{MIRD}
E.~{Hadad}, F.~{Heese}, P.~{Vary}, and S.~{Gannot}, ``Multichannel audio
  database in various acoustic environments,'' in \emph{Int. Workshop Acoust.
  Signal Enhance. (IWAENC)}, Sep. 2014, pp. 31--317.

\bibitem{TIMIT}
J.~S. Garofolo, L.~F. Lamel, W.~M. Fisher, J.~G. Fiscus, and D.~S. Pallett,
  ``{D}{A}{R}{P}{A} {T}{I}{M}{I}{T} acoustic-phonetic continous speech corpus
  {C}{D}-{R}{O}{M},'' 1993.

\bibitem{bss}
E.~Vincent, R.~Gribonval, and C.~F{\'e}votte, ``Performance measurement in
  blind audio source separation,'' \emph{IEEE Trans. Audio, Speech, Lang.
  Process.}, vol.~14, no.~4, pp. 1462--1469, 2006.

\end{thebibliography}

\end{document}